\documentclass[prl,aps,showpacs,twocolumn,superscriptaddress]{revtex4}
\usepackage{graphicx,amssymb,amsmath,color,psfrag}
\usepackage{amsthm}
\usepackage{amsfonts}
\usepackage{algorithmic}
\usepackage{enumerate}
\usepackage{latexsym}
\makeatletter

\newcommand{\Rmnum}[1]{\expandafter\@slowromancap\romannumeral #1@}
\makeatother

\begin{document}

\title{Electronic band gaps and transport properties in periodically alternating
mono- and bi-layer graphene superlattices}
\author{Xiong Fan}
\affiliation{Department of Physics, Beijing Normal University, Beijing 100875, China}
\author{Wenjun Huang}
\affiliation{Department of Physics, Beijing Normal University, Beijing 100875, China}
\author{Tianxing Ma}
\email{txma@bnu.edu.cn}
\affiliation{Department of Physics, Beijing Normal University, Beijing 100875, China}
\affiliation{Beijing Computational Science Research Center, Beijing 100084, China}
\author{Li-Gang Wang}
\email{sxwlg@yahoo.com}
\affiliation{Department of Physics, Zhejiang University, Hangzhou 310027, China}
\affiliation{Beijing Computational Science Research Centre, Beijing 100084, China}

\author{Hai-Qing Lin}
\affiliation{Beijing Computational Science Research Centre,
Beijing 100084, China}

\begin{abstract}
We investigate the electronic band structure and transport properties of periodically alternating mono- and bi-layer graphene superlattices (MBLG SLs). In such MBLG SLs, there exists a zero-averaged wave vector (zero-$\overline{k}$)  gap that is insensitive to the lattice constant. This zero-$\overline{k}$ gap can be controlled by changing both the ratio of the potential widths and the interlayer coupling coefficient of the bilayer graphene. We also show that there exist extra Dirac points; the conditions for these extra Dirac points are presented analytically. Lastly, we demonstrate that the electronic transport properties and the energy gap of the first two bands in MBLG SLs are tunable through adjustment of the interlayer coupling and the width ratio of the periodic mono- and bi-layer graphene.
\end{abstract}
\maketitle

Since it was first successfully fabricated in experiment approximately ten years ago \cite{Novoselov2004}, graphene has become an important research topic in condensed matter physics and material science. Indeed, because of its unique
characteristics, graphene has the potential to take the place of Si-based semiconductors in future applications \cite{Novoselov2004,Novoselov2006,AHCastro2009,Ma2010,Peres2010,Bolotin2008,Novoselov2005,Purewal,Lee2015,Gregersen2015}.
In recent years,  many important properties of graphene have been explored, such as chiral tunnelling \cite{Novoselov2006}, the giant carrier mobilities of this material \cite{Bolotin2008}, the unusual integer quantum Hall effect \cite{Novoselov2005,Purewal}, and the edge-dependant spectra of graphene nanoribbons \cite{Brey2006a,Nakada}. In particular, for graphene-based superlattices (SLs), new Dirac points (DPs) and the zero-averaged wave vector (zero-$\overline{k}$) gap have been observed \cite{MBarb2010,LBrey2009,LGWang2010,TXMa2012,changan2013}. The electronic properties of monolayer graphene (MLG) and bilayer graphene (BLG) SLs of various sequences have been studied using the transfer
matrix method \cite{Bai2010,LGWang2010,TXMa2012,changan2013,PLzhao2011,Gong2012,XXGuo2011,ZhRzh2012}.
In addition, structures with biased potentials \cite{Michael2009,johan2007} and heterostructures \cite{johan2007,Giannazzo2012,Ando2010,Yu2014} have been investigated. However, few studies have discussed the properties of periodic mono- and bi-layer graphene (MBLG) SLs, in which the MLG is decoupled from the BLG.

In this paper, we study the electronic band gaps and transport properties of MBLG SLs. We demonstrate that a zero-$\overline{k}$ gap and extra DPs exist under certain conditions. The zero-$\overline{k}$ gap opens and closes periodically, while the number of extra DPs increases with increased lattice constant $\Lambda$ values. The relationship between the zero-$\overline{k}$ gap and the interlayer coupling $t^{\prime }_{B}$ for MBLG SLs with periodic sequences is revealed. Furthermore, the effects of the \textit{average interlayer coupling} $\overline{t^\prime}$, which is dependant on $t^{\prime }_{B}$ and the ratio of the MLG and BLG widths ($w_A$ and $w_B$, respectively), on the electronic band structures of MBLG SLs are explored. This provides a tunable method of controlling electronic conductance using MLG and BLG by varying the $t^{\prime }_{B}$ values (note that, for MLG, the interlayer coupling is zero).

\begin{figure}[t b]
\centering
\includegraphics[scale=0.4]{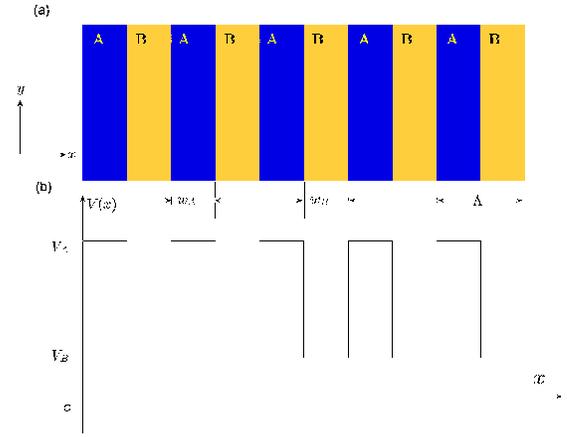}
\caption{(Color online)
{(a) Schematic representation of mono- and bi-layer graphene (MBLG) superlattice (SL) periodically aligned in $x$ direction. A and B denote monolayer (MLG) and bilayer graphene (BLG), respectively. $w_{A(B)}$ is the width of the A (B) region and $\Lambda=w_A+w_B$ is the lattice constant. Note that, in the MLG regions, there is no interlayer coupling and $t^{\prime}_j=0$ in Eq. (\ref{H}). In the BLG regions, the non-zero $t^{\prime}_j$ defined in Eq. (\ref{H}) occurs. (b) Profiles of periodic potentials applied on graphene SL. $V_{A(B)}$ is the square potential applied on the A (B) region.}}
\label{Fig:fig1}
\end{figure}

We begin with the electronic structure of BLG with the energy and wave vector close to the $K$ point, such that the one-particle Hamiltonian for the BLG is given by \cite{Johan2006,Edward2006,Snyman2007,Michael2009,johan2007,Ando2010,Yu2014,MZarenia2012}
\begin{equation}
H_{j}=\left(
\begin{array}{cccc}
V_{j}(x) & \pi  & t_{j}^{\prime } & 0 \\
\pi ^{\dag } & V_{j}(x) & 0 & 0 \\
t_{j}^{\prime } & 0 & V_{j}(x) & \pi ^{\dag } \\
0 & 0 & \pi  & V_{j}(x)
\end{array}
\right) .  \label{H}
\end{equation}
Here, $V_{j}(x)$ is the electrostatic potential applied on the material, which is constant within each potential barrier or well. Further, $\pi =-i\hbar \upsilon _{F}[\frac{\partial }{\partial x}-i\frac{\partial }{\partial y}]$ and $\pi ^{\dag}=-i\hbar \upsilon _{F}[\frac{\partial }{\partial x}+i\frac{\partial }{
\partial y}]$ are the momentum operators,
where $\upsilon_{F}\approx 10^{6}$ m/s is the Fermi velocity and $t_{j}^{\prime }$ indicates the interlayer coupling in the BLG regions. It should be noted that $t_{j}^{\prime }$ can be tuned by adjusting the interlayer distance \cite{Long2015}. The wave function is expressed by four-component pseudospinors $\Phi={\left(\widetilde{\varphi }_{1},\widetilde{\varphi }_{2},\widetilde{\varphi }_{3},\widetilde{\varphi }_{4}\right) }^{T}$. Note that, when $t_{j}^{\prime }=0$, Eq. (\ref{H}) reduces to the MLG case. As a result of the translation invariance in the $y$ direction, the wave function can be rewritten as $\widetilde{\varphi }_{m}=\varphi _{m}e^{ik_{y}y}$, $m=1$,...,$4$. By solving this eigenequation, the wave functions at any two positions $x$ and $x+\Delta x$ inside the $j$th potential can be related by a transfer matrix \cite{changan2013}
\begin{equation}
M_{j}=
\begin{pmatrix}
M_{+} & 0 \\
0 & M_{-}
\end{pmatrix}
,
\end{equation}
and
\begin{equation}
M_{\pm }=
\begin{pmatrix}
\frac{\cos (q_{j}\Delta x\mp \Omega _{j})}{\cos \Omega _{j}} & i\frac{k_{j}}{
q_{j}}\sin (q_{j}\Delta x) \\
i\frac{k_{j}^{\prime }\sin (q_{j}\Delta x)}{k_{j}\cos \Omega _{j}} & \frac{
\cos (q_{j}\Delta x\pm \Omega _{j})}{\cos \Omega _{j}})\label{eq3}
\end{pmatrix}
,
\end{equation}
{where $k_{j}$=$(E-V_{j})/\hbar \upsilon _{F}$, $\Delta x$ is the interval of any two positions inside the $j$th potential, $q_{j}=$ sign($k_{j}$)$\sqrt{k_{j}^{2}-k_{y}^{2}-t_{j}^{\prime }k_{j}}$ for $k_{j}^{2}-k_{y}^{2}-t_{j}^{\prime }k_{j}>0$, otherwise $q_{j}=i\sqrt{|k_{j}^{2}-k_{y}^{2}-t_{j}^{\prime }k_{j}|}$, $t_{j}^{\prime }\rightarrow t_{j}^{\prime }/\hbar \upsilon _{F}$, $k_{j}^{\prime 2}$=$k_{y}^{2}+q_{j}^{2}$, and $\Omega _{j}$ = arcsin$(k_{y}/k_{j}^{\prime })$.}

If $t_{j}^{\prime }=0$, i.e., the BLG is decoupled as MLG, the transfer matrix becomes
\begin{equation}
M_{\pm }=
\begin{pmatrix}
\frac{\cos (q_{j}\Delta x\mp \Omega _{j})}{\cos \Omega _{j}} & i\frac{\sin
(q_{j}\Delta x)}{\cos \Omega _{j}} \\
i\frac{\sin (q_{j}\Delta x)}{\cos \Omega _{j}} & \frac{\cos (q_{j}\Delta
x\pm \Omega _{j})}{\cos \Omega _{j}})
\end{pmatrix}
.
\end{equation}
Compared with the BLG case, for MLG with $t^{\prime}_j=0$, the wave vector $k_{j}^{\prime }$ inside the $j$th potential reduces to $k_{j}$. The other parameters have the same forms in both cases.

Using the boundary conditions, the transmission coefficient $t=t(E,k_{y})$ can be expressed as
\begin{equation}
t=\frac{2\cos \Omega _{0}}{(x_{22}e^{-i\Omega _{0}}+x_{11}e^{i\Omega
_{e}})-x_{12}e^{i(\Omega _{e}-\Omega _{0})}-x_{21}},
\end{equation}
where $x_{i,j}\,(i,j=1,2)$ are elements of the total transfer matrix $X_{N}$=$\prod_{j=1}^{N}M_{j}$, with $N$ being the total number of potential barriers and wells. The transmissivity is $T(E,k_{y})=|t(E,k_{y})|^2$.

Let us consider a periodic sequence of MBLG SLs, which is perhaps the simplest sequence (see Fig. \ref{Fig:fig1}). Here, the width and the applied potential of the MLG on each unit are $w_{A}$ and $V_{A}$, respectively, and the corresponding BLG parameters on each unit are $w_{B}$ and $V_{B}$, respectively. According to Bloch's theorem, for an infinite periodic structure $(AB)^{N}$, the electronic dispersion at any incident angle follows the relation
\begin{eqnarray}
\cos{[\beta _{x}\Lambda ]}&=&\frac{1}{4}Tr[M_{A}M_{B}], \notag \\
&=&\cos {(q_{A}w_{A}+q_{B}w_{B})}+\sin {(q_{A}w_{A})}\sin {(q_{B}w_{B})} \notag \\
&&\times \frac{2k_{B}(q_{A}q_{B}+k_{y}^{2})-k_{A}(k_{B}^{2}+k_{B}^{\prime 2})}{2k_{B}q_{A}q_{B}},\label{COSBL}
\end{eqnarray}
where $\Lambda =w_{A}+w_{B}$. Using $|\cos{[\beta _{x}\Lambda ]}| \leq 1$, we can find the real solution of $\beta_{x}$ for passing bands. Otherwise, the non-existence of real $\beta _{x}$ indicates a band gap.
\begin{figure}[tbp]
\includegraphics[scale=0.36]{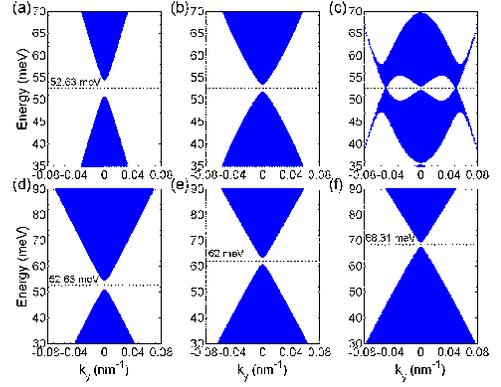} \centering
\caption{(Color online) Electronic band structures for different $\Lambda$ with $w_{A}=w_{B}=$ (a) 15,
(b) 30, and (c) 60 nm, and electronic band structures for different potential width ratios $w_{A}/w_{B}=$ (d) 1, (e) 3/2, and (f) 2/1 under fixed $w_{B}=15$ nm. The dotted lines denote the centre position of the zero-$\overline{k}$ gap. The other parameters are $V_{A}=100$ meV, $V_{B}=0$ meV, and $t^{\prime }_{B}=10$ meV.}
\label{Fig:fig2}
\end{figure}

Figure \ref{Fig:fig2} demonstrates the properties of electronic band gaps under different lattice parameters, indicating that there is a zero-$\overline{k}$ gap in such periodic MBLG SLs. From Figs. \ref{Fig:fig2}(a)--\ref{Fig:fig2}(c), it is apparent that the location of the zero-$\overline{k}$ gap is independent of $\Lambda$. In this case, it is positioned at approximately $52.63$ meV for a variety of $\Lambda$. Under the same lattice parameters, the location of the zero-$\overline{k}$ gap in periodic MBLG SLs is higher and lower than those of MLG \cite{TXMa2012} and BLG \cite{changan2013} SLs, respectively. Extra DPs may appear at $k_{y}\neq 0$ for larger $\Lambda$, as shown in Fig. \ref{Fig:fig2}(c); this will be discussed below. From Figs. \ref{Fig:fig2}(d)--\ref{Fig:fig2}(f), the location of the zero-$\overline{k}$ gap varies with changes in the $w_{A}/w_{B}$ ratio. For example, the zero-$\overline{k}$ gaps are positioned at energy $E=52.63$, $62.00$, and $68.31$ meV for $w_{A}/w_{B}=1$, $3/2$, and $2/1$, respectively.

The location of the zero-$\overline{k}$ gap is determined by $\overline{k}=\sum_{j=1}^{N}k_{j}w_{j}/\sum_{j=1}^{N}w_{j}=0$ \cite{TXMa2012}. Note that $k_{j}$ should be replaced by $k_{j}^{\prime }$ for BLG. When the number of the MLG regions is equal to that of the BLG regions, the $E$ corresponding to $\overline{k}=0$ can be easily found. Initially,
\begin{eqnarray}
 &&E=\frac{2(V_{A}w_{A}^{2}-V_{B}w_{B}^{2})-t^{\prime }_{B}w_{B}^{2}}{2(w_{A}^{2}-w_{B}^{2})} \notag \\
  &&-\frac{\sqrt{w_{A}^{2}w_{B}^{2}(2V_{A}-2V_{B}-t^{\prime}_{B})^2+t^{\prime2}_{B}w_{B}^{2}(w_{B}-w_{A})^{2}}}{2(w_{A}^{2}-w_{B}^{2})}.
\label{Eq:eq7}
\end{eqnarray}
Then, for $w_{A}=w_{B}$, Eq. (\ref{Eq:eq7}) reduces to
\begin{equation}
E=\frac{V_{A}^{2}-V_{B}^{2}-t^{\prime }_{B}V_{B}}{2V_{A}-2V_{B}-t^{\prime }_{B}}.
\label{Eq:eq8}
\end{equation}
The locations of the zero-$\overline{k}$ gaps in Fig. \ref{Fig:fig2} are in agreement with Eqs. (\ref{Eq:eq7}) and (\ref{Eq:eq8}).

We now consider a method to determine the locations of the extra DPs, which should obey the equation
\begin{equation}
\cos{[\beta _{x}\Lambda ]}=1.  \label{Eq:eq9}
\end{equation}
Using Eq. (\ref{COSBL}), when $q_{A}w_{A}=-q_{B}w_{B}=m\pi $ ($m$ is a positive integer), Eq. (\ref{Eq:eq9}) above is satisfied. For $w_{A}=w_{B}=w$, the zero-$\overline{k}$ gap will close at the normal
incident angle ($k_{y}=0$) with $w$ satisfying the condition
\begin{equation}
w_{m}=\frac{m\pi \hbar \upsilon _{F}(2V_{A}-2V_{B}-t^{\prime }_{B})}{
(V_{B}-V_{A})(V_{B}-V_{A}+t^{\prime }_{B})}.\ (m=1,2,3\dots)
\label{Eq:eq10}
\end{equation}
At oblique incidences ($k_{y}\neq 0$) and when $q_{A}w_{A}=-q_{B}w_{B}=m\pi $, the extra DPs appear. They are located at
\begin{equation}
k_{y,m}=\pm \sqrt{\frac{(V_{A}-V_{B})^{2}(t^{\prime }_{B}-V_{A}+V_{B})^{2}}{
\hbar ^{2}\upsilon _{F}^{2}(2V_{A}-2V_{B}-t^{\prime }_{B})^2}-\left( \frac{m\pi }{w
}\right) ^{2}}.  \label{Eq:eq11}
\end{equation}
The number of extra DPs can be obtained from Eq. (\ref{Eq:eq11}) using the limiting condition: $k_{y}\leq |E/\hbar\upsilon _{F}|$.
\begin{figure}[t b p]
\includegraphics[scale=0.44]{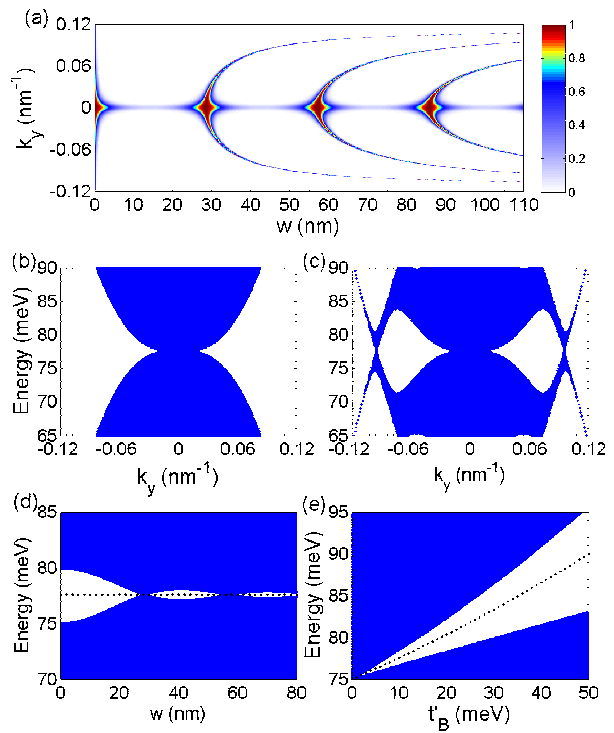}
\centering
\caption{(Color online) (a) Transmission probabilities of zero-$\overline{k}$ gap centre in finite periodic potential structure $(AB)^{25}$. Electronic band structures for $w_{A}=w_{B}=$ (b) 28.56 and (c) 57.12 nm. (d) Band-gap structure dependence on $w$ when $w_{A}=w_{B}=w$ with $k_{y}=0$. (e) Effect of $t^{\prime }_{B}$ on band-gap structure with $w_{A}=w_{B}=20$ nm. The other parameters are $V_{A}=150$ meV, $V_{B}=0$ meV, and $t^{\prime }_{B}=10$ meV. }
\label{Fig:fig3}
\end{figure}

Extra DPs exist in the band structures of periodic MBLG SLs, which have already been shown in Fig. \ref{Fig:fig2}(c). In Fig. \ref{Fig:fig3}(a), the transmission probabilities of the centre of the zero-$\overline{k}$ gap are plotted in order to find the extra DPs. In MLG SLs with periodic sequences, the charge carriers have perfect transmission at normal incidence and a DP is always positioned at the centre of the gap \cite{LGWang2010}. However, a different scenario occurs in the case of MBLG SLs with periodic sequences. As illustrated in Fig. \ref{Fig:fig3}(a), the zero-$\overline{k}$ gap closes with a fixed period and extra paired DPs occur as $w$ increases; this is characterized by the large transmission probability in Fig. \ref{Fig:fig3}(a). In addition, it is important to notice that the DP does not exist at the normal incidence, and that it always has large reflection unless the gap is closed. From Eq. (\ref{Eq:eq10}), the gap-closing period is $28.56$ nm for the parameters given in Fig. \ref{Fig:fig3}(a). Besides, extra DPs do not exist unless $\Lambda$ is larger than one period, as can be seen in Figs. \ref{Fig:fig3}(a)--\ref{Fig:fig3}(c). Fig. \ref{Fig:fig3}(d) shows that the centre position of the zero-$\overline{k}$ gap is independent of $\Lambda$. The effect of the BLG's $t^{\prime}_{B}$ on the band structure is shown in Fig. \ref{Fig:fig3}(e). It is apparent that the width of the zero-$\overline{k}$ gap increases significantly as $t^{\prime }_{B}$ increases. The dotted dark line denotes the centre position of the zero-$\overline{k}$ gap and satisfies a nonlinear relation [see Eq. (\ref{Eq:eq8})]. This differs from the linear relation obtained in the case of BLG SLs \cite{changan2013}.
\begin{figure}[htbp]
\includegraphics[scale=0.4]{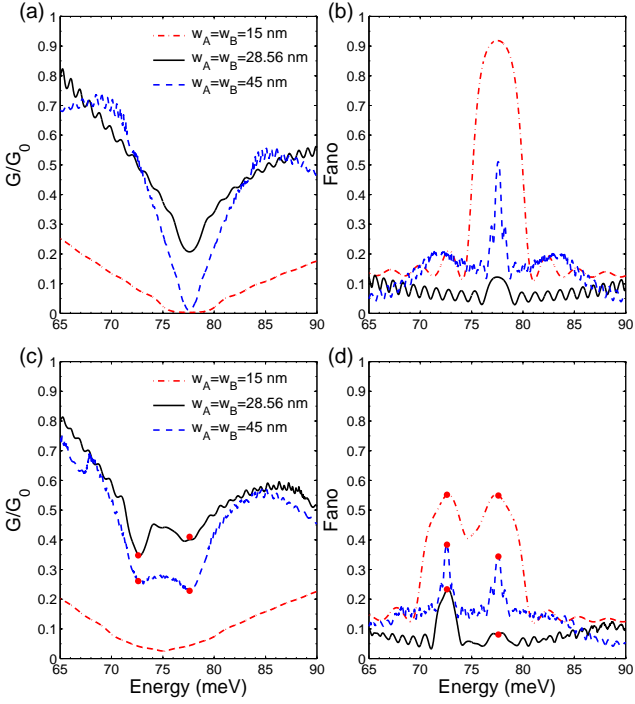}
\centering
\caption{(Color online) (a) Conductance and (b) Fano factor vs Fermi energy focusing on one-mode wave function in finite periodic sequence $(AB)^{25}$. Averages of (c) conductance and (d) Fano factor vs Fermi energy considering two-mode wave function in finite periodic sequence $(AB)^{25}$. The other parameters are identical to those in Fig. \ref{Fig:fig3}.}
\label{Fig:fig4}
\end{figure}

We next consider the total conductance and Fano factor (the ratio between the shot noise power and current) \cite{Datta1995,tworzydlo,GF} for different $\Lambda$. In Fig. \ref{Fig:fig4}(a), the conductance of the zero-$\overline{k}$ gap exhibits different values for different $\Lambda$. For fixed $\Lambda$, where $w_{A}=w_{B}=15$ nm, the conductance is approximately zero when the zero-$\overline{k}$ gap opens. Specifically, a pair of extra DPs exist at the centre position of the zero-$\overline{k}$ gap for  $w_{A}=w_{B}=45$ nm [see Fig. \ref{Fig:fig3}(a)]. In this case, the conductance of this position is slightly higher than zero and the angular-average conductance curve forms a linear-like cone around the extra DPs. There is no unique DP at the zero-$\overline{k}$ gap in MBLG SLs with periodic sequences, as is the case for MLG SL \cite{TXMa2012,PLzhao2011,XXGuo2011}. Therefore, the Fano factor at the centre position of the zero-$\overline{k}$ gap can be removed from $1/3$.

All the previous results were obtained by focusing on the one-mode wave function with longitudinal wave vector $q_j$. Experimentally, this scenario can be realized using, for example, an adatom to create a wave function with a certain longitudinal wave vector. According to the result presented in Ref. \onlinecite{duppen2013}, the two-mode wave function with no mixing can be considered as the primary contributor to the conductance or the Fano factor [in Ref. \onlinecite{duppen2013}, see the values of $T_{+}^{+}$ and $T_{-}^{-}$ and compare them with $T_{+}^{-}$ and $T_{-}^{+}$]. Here, we ignore other complex transport processes that mix different wave function modes, and instead consider the contribution of the wave function with longitudinal wave vector $q_{j}^{\prime}=$ sign($k_{j}$)$\sqrt{k_{j}^{2}-k_{y}^{2}+t_{j}^{\prime }k_{j}}$ for $k_{j}^{2}-k_{y}^{2}+t_{j}^{\prime }k_{j}>0$, otherwise $q_{j}^{\prime}=i\sqrt{|k_{j}^{2}-k_{y}^{2}+t_{j}^{\prime }k_{j}|}$. The averaged conductance and Fano factor produced by both modes of the wave functions with $q_j$ and $q'_j$ are plotted in Figs. \ref{Fig:fig4}(c) and \ref{Fig:fig4}(d), respectively. In these figures, the conductance curves exhibit several dips, whereas the Fano factor curves exhibit several small peaks, all of which are indicated by red dots. The groups of dips or peaks are located at $72.6$ and $77.6$ meV and are, therefore, separated; these separations are primarily caused by the zero-$\overline{k}$ gaps of the wave functions with $q'_j$ and $q_j$. For the case of $w_A=w_B=15$ nm, the average of the conductance at approximately $77.6$ meV is increased compared with the zero conductance in Fig. \ref{Fig:fig4}(a), and the position of the conductance minimum is changed from $77.6$ to $75.0$ meV. Based on the above analysis, it may be concluded that the result obtained by focusing on the one-mode wave function is meaningful as a reflection of the transport properties of normal scenarios involving the two-mode wave function.

\begin{figure}[b t p]
\includegraphics[scale=0.18]{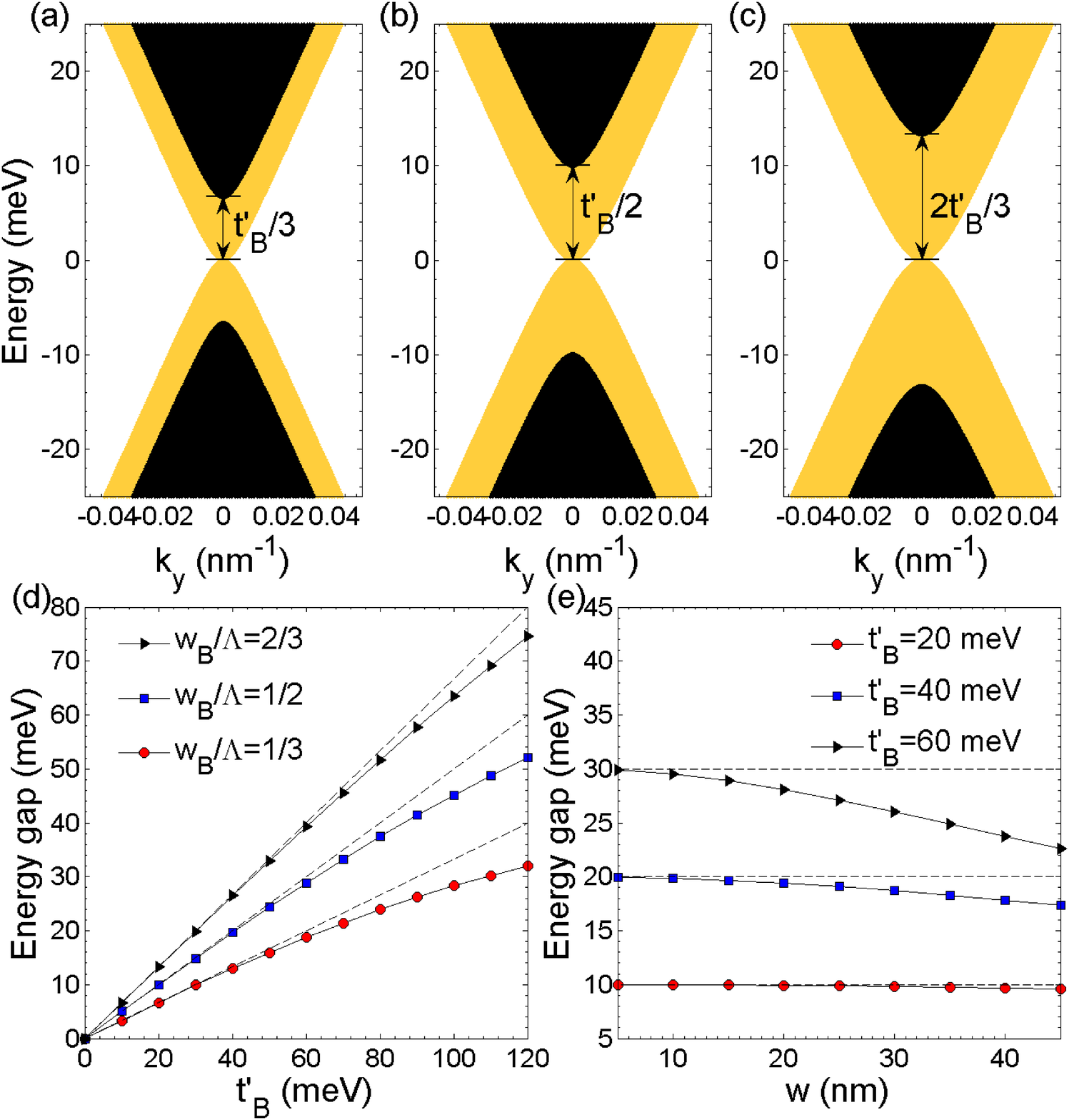}
\centering
\caption{(Color online) Electronic band structures for $w_{A}/w_{B}=$ (a) 2, (b) 1, and (c) 1/2 with fixed $w_{B}=15$ nm. The other parameters are $V_{A}=V_{B}=0$ meV and $t^{\prime }_{B}=20$ meV. (d) Energy gap vs $t^{\prime }_{B}$. (e) Energy gap vs $w$ for $w_{A}=w_{B}=w$. The double arrows in (a)--(c) denote the energy gap widths of the two modes and the dashed lines in (d) and (e) denote the relation given in Eq. (\ref{Eq:eq12}).}
\label{Fig:fig5}
\end{figure}
Finally, we consider another interesting effect. In the absence of an applied potential on the periodic MBLG SLs, the energy gap of the first two bands can be regularly tuned by controlling the $w_A/w_B$ ratio. The yellow regions in Figs. \ref{Fig:fig5}(a)--\ref{Fig:fig5}(c) are occupied by electrons of the propagating mode, whereas the dark regions are occupied by electrons of both the propagating and evanescent modes. As illustrated in these figures, the gap width of the first two bands changes from $t_{B}^{\prime}/3$, $t_{B}^{\prime}/2$, to $2t_{B}^{\prime}/3$ for $w_A/w_B = 2$, $1$, and $1/2$, respectively, under the given parameters. Further, the gap width is proportional to the $w_{B}/(w_{A}+w_{B})$ ratio and roughly obeys
\begin{equation}
E_{g}=\frac{t^{\prime }_{B}w_{B}}{w_{A}+w_{B}}.
\label{Eq:eq12}
\end{equation}
Figs. \ref{Fig:fig5}(d) and \ref{Fig:fig5}(e) show the correct range of Eq. (\ref{Eq:eq12}). From Fig. \ref{Fig:fig5}(d), it is apparent that Eq. (\ref{Eq:eq12}) accurately describes the regularity when $t^{\prime}_{B}$ is smaller than approximately $50$ meV in the case of $w_B=15$ nm, and the deviation is larger for larger $t^{\prime}_{B}$. $\Lambda$ also affects the correctness of the above relation, as shown in Fig. \ref{Fig:fig5}(e). It is concluded that the actual gap width is smaller than the approximated result given by Eq. (\ref{Eq:eq12}), which indicates that, as $t^{\prime }_{B}$ or $w$ increases, the effect of the MLG becomes stronger than that of the BLG. Based on the similarity between the band structures in Fig. \ref{Fig:fig5} and those of the BLG, we may define $E_g$ as $\overline{t^\prime}$. As $\overline{t^\prime}$ increases, the band structure becomes more similar to the parabolic band structure of BLG. 
For a periodic structure with alternating interlayer couplings in total two-layer graphene, Eq. (\ref{Eq:eq12}) (which concerns the gap width) can be modified to
\begin{equation}
E_{g}=\frac{t_{A}'w_{A}+t_{B}'w_{B}}{w_{A}+w_{B}},
\end{equation}
where $t_{A}'$ is the interlayer coupling in the A regions, which are also BLG.
The $\overline{t^\prime}$ in MBLG SLs corresponds to the border between the propagating and evanescent wave modes, and this BLG border has been shown in Duppen and Peeters' interesting work (see Ref. \onlinecite{duppen2013}). Furthermore, this border is tunable through adjustment of $t_{A}'$ and $t_{B}'$  and $w_{A}/w_{B}$.
Therefore, this may constitute a new method of controlling the electronic conductance.

In summary, we have studied the electronic band structures and transport properties of mono- and bi-layer graphene (MBLG) superlattices (SLs) with periodic sequences. The zero-$\overline{k}$ gap is robust against the lattice constant $\Lambda = w_A + w_B$, but sensitive to both the ratio of the potential widths $w_{A}/w_{B}$ and the interlayer coupling $t^{\prime }_{B}$. The analytical condition determining the locations of extra Dirac points (DPs) ($k_{y}\ne 0$) was presented and it was shown that the number of DPs increases as $\Lambda$ increases. Moreover, the conductance of the charge carriers in periodic MBLG SLs was investigated, and it was shown that the angular-average conductance curve forms a linear-like cone around the extra DPs, similar to that observed in the case of MLG SLs \cite{TXMa2012}. If no potential barriers are applied, the energy gap width of the first two bands or the average interlayer coupling $\overline{t^\prime}$ is tunable through adjustment of $t^{\prime }_{B}$ and $w_{A}/w_{B}$. This finding is important as regards research on interlayer coupling in two-layer graphene, and may constitute a new method of controlling the electronic conductance. Further related results will be reported in the near future. The findings of this and future studies will be useful for future research and applications concerning MBLG SLs.

T. Ma thanks CAEP for partial financial support. This work is supported by NSFCs (Grant. Nos. 11274275 and 11374034), and the National Basic Research Program of China (Grant Nos. 2011CBA00108 and 2012CB921602). We also acknowledge support from the Fundamental Research Funds for the Center Universities under Nos. 2015FZA3002 (L.-G. Wang) and 2014KJJCB26 (T. Ma).


\begin{thebibliography}{99}
\bibitem{GF} G and F are calculated by $G=G_0\int_{0}^{\pi/2}T\cos \theta_{0}d\theta_{0}$ and $F=\int_{-\pi/2}^{\pi/2}(1-T)T\cos \theta_{0}d\theta_{0}/\int_{-\pi/2}^{\pi/2}T\cos \theta_{0}d\theta_{0}$, where $G_0=2e^2mv_{F}L_{y}/\hbar^2$ ($L_{y}$ denotes the length of graphene stripe in the y direction).
\end{thebibliography}
\end{document}